\pdfoutput=1

\documentclass[11pt,prd,aps,amssymb,amsmath,tightenlines,showpacs]{revtex4}
\usepackage{graphicx}

\usepackage{hyperref}   
\hypersetup{colorlinks=true,linkcolor=blue, citecolor=blue}

\begin{document}

\title{Effects of complex parameters on classical trajectories of\\ Hamiltonian systems}

\author{Asiri Nanayakkara}\email{asiri@ifs.ac.lk}
\author{ Thilagarajah Mathanaranjan$^*$}\email{mathan@jfn.ac.lk}

\affiliation{$^*$Institute of Fundamental Studies Hanthana Road, Kandy, Sri Lanka\\
$^\dagger$ Department of Mathematics, University of Jaffna, Sri Lanka}

\date{\today}

\begin{abstract}
Anderson et al have shown that for complex energies, the classical
trajectories of \textit{real }quartic potentials are closed and periodic
only on a discrete set of eigencurves. Moreover, recently it was revealed
that, when time is complex $t$ $(t=t_{r}e^{i\theta _{\tau }}),$ certain real
hermitian systems possess close periodic trajectories only for a discrete
set of values of $\theta _{\tau }$. On the other hand it is generally true
that even for real energies, classical trajectories of non $\mathcal{PT}$- symmetric
Hamiltonians with complex parameters are mostly non-periodic and open. In
this paper we show that for given real energy, the classical trajectories of 
\textit{complex} quartic Hamiltonians $H=p^{2}+ax^{4}+bx^{k}$, (where $a$ is
real, $b$ is complex and $k=1$ $or$ $2$) are closed and periodic only for a
discrete set of parameter curves in the complex $b$-plane. It was further
found that given complex parameter $b$, the classical trajectories are
periodic for a discrete set of real energies (i.e. classical energy get
discretized or quantized by imposing the condition that trajectories are
periodic and closed). Moreover, we show that for real and positive energies
(continuous), the classical trajectories of \textit{complex} Hamiltonian $%
H=p^{2}+%
\mu
x^{4},$ ($%
\mu
=\mu _{r}e^{i\theta })$ are periodic when \ $\theta =4 tan^{-1}[(n/(2m+n))]$
for $\forall $ $n$ and $m\in 
\mathbb{Z}
$.

\end{abstract}


\maketitle

\section{\textbf{Introduction}}\label{sec:1}

In recent years, classical behavior of non-Hermitian Hamiltonian systems\cite{R1,R2,R3,R4,R5} as well as classical motion of Hermitian systems for complex energies\cite{R6,R7,R8,R9,R10,R11,R12,R13,R14,R15,R16} have attracted much interest. Investigation of classical mechanics in the complex domain is useful for understanding various classical and quantum
mechanical phenomena such as barrier tunneling, dynamical tunneling\cite{R17,R18}, classical and quantum chaos\cite{R7,R19}, quantum correspondence principle,complex forms of uncertainty relations and the semiclassical limit of complex quantum field theories.

Since the earlier work on classical motion of non-Hermitian systems \cite{R6,R7},
there have been several interesting results found on the various aspects of
the subject\cite{R8,R9,R10,R11,R12,R13,R14,R15,R16}. Numerical and analytical investigations have revealed
that when energies are real, classical trajectories of complex $\mathcal{PT}$-symmetric
non Hermitian systems are closed and periodic. However, when energies of
these systems are complex, the periodic trajectories usually become non
periodic and open\cite{R15,R19}. Recently it was shown that even though most of
the trajectories corresponding to complex energies are open and non
periodic, for some systems, there are special discrete sets of curves in the
complex-energy plane for which the trajectories are periodic\cite{R20}. On the
other hand, in non-Hermitian and non $\mathcal{PT}$-symmetric Hamiltonian systems, even
for real energies, almost all trajectories except a few are non-periodic and
open. It was also shown recently that when time is taken as a complex
quantity with a specific fixed phase angle or as a specific complex
function, non periodic trajectories of 1-D Hamiltonian systems become
periodic and closed\cite{R21}.

In this paper we investigate the classical trajectories from a different
point of view. Here we examine clasical behavior of the complex Hamiltonian $%
H=p^{2}+ax^{4}+bx^{k}$, (where $k=1$, $2$, and $a$ is real such that $H$ is
not $\mathcal{PT}$-symmetric) for complex parameter $b$ and real energy $E$. Outline of
the paper is as follows. In the section \ref{sec:2}, analytic expressions for complex
trajectories are derived.Expressions for periods of the periodic trajectories as well as time taken by unbounded trajectories to escape to
infinity are found in terms of $b$ and energy $E$. \ We will show that for
given real energy, the classical trajectories of the above quartic
Hamiltonian are open except for a discrete set of parameter values in the
complex $b$-plane. In section \ref{sec:3} we study how trajectories behave when energy
is real and $b$ is a fixed complex parameter. The classical trajectories of
complex Hamiltonian $H=p^{2}+%
\mu
x^{4},$ ($%
\mu
=re^{i\theta })$ is investigated for real energies in section \ref{sec:4} and
concluding remarks are given in section \ref{sec:5}.

\section{\protect \protect\textbf{Classical trajectories of 
}$H=p^{2}+ax^{4}+bx^{k}$}\label{sec:2}

In this section first we study in detail the classical motion of the complex
quartic anharmonic oscillator. We assume that the Hamiltonian has the form
\begin{equation}
H=p^{2}+ax^{4}+bx^{k} \label{eq:1}
\end{equation}%
where $a$ is a real positive constant, $b$ is a complex constant and $k=1$
or $2$. First we derive expressions for $x(t)$ and the period for the above
Hamiltonian. When it is needed, value of $k$ is chosen as $1$ or $2$.
Throughout this paper mass of the particle is taken as half $(i.e.2m=1).$
The equation of motion is
\begin{equation}
\frac{dx}{dt}=p=2\sqrt{E-ax^{4}-bx^{k}}\label{eq:2}
\end{equation}
The turning points of this system are taken as $x_{0},x_{1},x_{2}$ and $%
x_{3} $ and by integrating equation(\ref{eq:2})we have
\begin{equation}
\int \frac{dx}{\sqrt{\left( x-x_{0}\right) (x-x_{1})(x-x_{2})(x-x_{3})}}=2%
\sqrt{a}e^{i\pi /2}t+c  \label{eq:3}
\end{equation}%
where $c$ is the constant of integration which depends on initial
conditions. The left-hand side of the above equation is an elliptic integral
of the first kind and hence equation (\ref{eq:3}) becomes
\begin{equation*}
\frac{2}{\sqrt{\left( x_{0}-x_{2}\right) \left( x_{1}-x_{3}\right) }}{\small %
F}\left( \sin ^{-1}\left[ \sqrt{\frac{\left( x-x_{1}\right) (x_{0}-x_{2})}{%
\left( x-x_{0}\right) (x_{1}-x_{2})}}\right] ,\frac{\left(
x_{1}-x_{2}\right) (x_{0}-x_{3})}{\left( x_{0}-x_{2}\right) (x_{1}-x_{3})}%
\right)
\end{equation*}
\begin{equation}
=2\sqrt{a}e^{i\pi /2}t+c   \label{eq:4}
\end{equation}
where $F$\ is an elliptic function. We invert the above equation in terms of
Jacobian elliptic function `$sn$' as
\begin{equation}
x\left( t\right) =\frac{x_{1}\left( x_{0}-x_{2}\right)
-x_{0}(x_{1}-x_{2})sn^{2}(u)}{\left( x_{0}-x_{2}\right)
-(x_{1}-x_{2})sn^{2}(u)}   \label{eq:5}
\end{equation}%
where $u=\sqrt{a\left( x_{0}-x_{2}\right) \left( x_{1}-x_{3}\right) }e^{i\pi
/2}t+\alpha $ and modulus $\kappa =\left[ \frac{\left( x_{1}-x_{2}\right)
(x_{0}-x_{3})}{\left( x_{0}-x_{2}\right) (x_{1}-x_{3})}\right] ^{1/2}$ and $%
\alpha $ is an arbitrary constant which is determined by the initial
conditions. Note that $x(t)$ in the above equation is still a solution of
(\ref{eq:3}), when $x_{0},x_{1},x_{2}$ and $x_{3}$ are cyclically changed (e.g. $%
x_{3}\rightarrow x_{0}\rightarrow x_{1}\rightarrow x_{2}\rightarrow x_{3}$).
In order to understand how the trajectories behave, we need to recognize the
periodic, bounded and unbounded properties of the function $x\left( t\right) 
$. First we find complementary modulus $\kappa ^{\prime }$ and complete
elliptic functions $K$ and $K^{\prime }$. They are defined by
\begin{equation}
\kappa ^{{\prime}^ 2}=1-\kappa ^{2}=\frac{\left( x_{0}-x_{1}\right)
(x_{2}-x_{3})}{\left( x_{0}-x_{2}\right) (x_{1}-x_{3})}    \label{eq:6}
\end{equation}
\begin{equation}
K=\overset{\frac{\pi }{2}}{\underset{0}{\int }}(1-\kappa ^{2}\sin ^{2}\left(
\phi \right) )^{-\frac{1}{2}}d\phi    \label{eq:7}
\end{equation}
\begin{equation}
 K^{\prime }=\overset{1}{\underset{0}{\int }}\left( 1-t^{2}\right) ^{-%
\frac{1}{2}}\left( 1-\kappa ^{{\prime}^2}t^{2}\right) ^{-\frac{1}{2}}dt.   \label{eq:8}
\end{equation}
$K$ and $K^{\prime }$ are evaluated directly from the above equations and
they are independent of phase angle $\theta $.

\begin{figure}[h]
\centering  
\includegraphics[width=7.5cm,height=6.0cm]{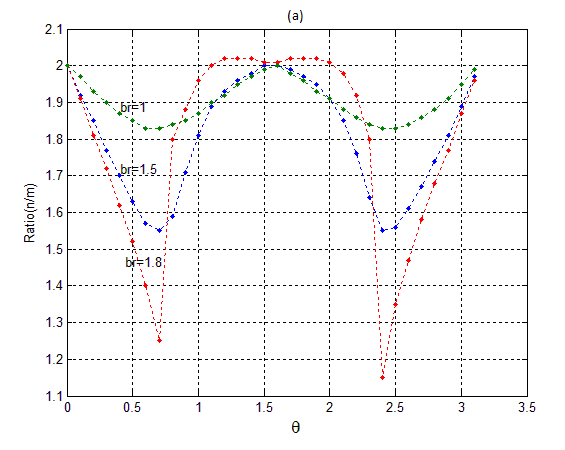}
\includegraphics[width=7.5cm,height=6.0cm]{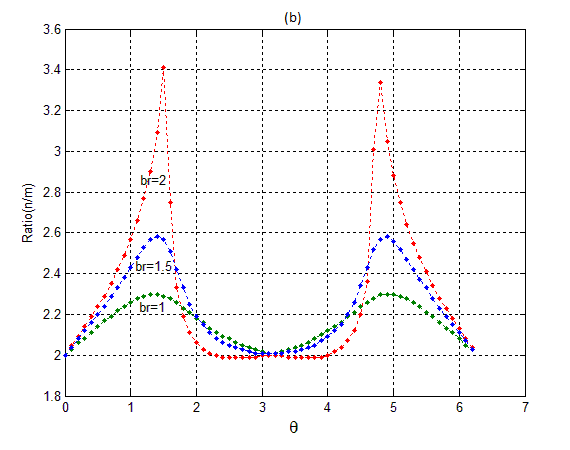} 
\caption{\small{By applying the condition that classical trajectories are
periodic, we obtain discrete values of $\theta $ for a fixed value of $b_{r}$%
. \textbf{(a)} shows how the ratio $r$ \ varies with complex phase angle $%
\theta $ of the potential $V(x)=x^{4}+b_{r}e^{i\theta }x\ $. In order to
have periodic trajectories, $r$ ($r\equiv \frac{n}{m}$) has to be rational
and hence only discrete values of $\theta $ satisfy the condition (13). Each
point in the graph represent such a value. \textbf{(b)} same as (a) but for
the potential $V(x)=x^{4}+b_{r}e^{i\theta }x^{2}$}}
\label{fig1}
\end{figure}

The trajectory $x(t)$ is given by $\ x\left( t\right) =\frac{x_{1}\left(
x_{0}-x_{2}\right) -x_{0}(x_{1}-x_{2})sn^{2}(u)}{\left( x_{0}-x_{2}\right)
-(x_{1}-x_{2})sn^{2}(u)}.$ The condition of trajectory become unbounded and
the particle escapes to infinity is
\begin{equation}
\left( x_{0}-x_{2}\right) -(x_{1}-x_{2})sn^{2}(u)=0 \bigskip   \label{eq:9}
\end{equation}
where $u=\sqrt{a\left( x_{0}-x_{2}\right) \left( x_{1}-x_{3}\right) 
}e^{i\pi /2}t+\alpha $ satisfied for some real positive $t$ and the time
taken for the particle to escape to $\infty $ is give by

\begin{equation}
T_{\infty }=\frac{\left( sn^{-1}(z_{0})-\alpha \right) {\small e}^{-i\pi /2}%
}{\sqrt{a\left( x_{0}-x_{2}\right) \left( x_{1}-x_{3}\right) }}   \label{eq:10}
\end{equation}
where $z_{0}=\sqrt{\frac{(x_{0}-x_{2})}{(x_{1}-x_{2})}}.$
`$sn$' is doubly periodic with period $4mK+2niK^{\prime }$ where $m$ and $n$
are integers. Therefore the condition for the trajectory to become periodic
and particle does not escape to infinity is
\begin{equation}
\sqrt{a\left( x_{0}-x_{2}\right) \left( x_{1}-x_{3}\right) }e^{i\pi
/2}t=4mK+2niK^{\prime }\ ;\ \ \ \ m,\ n\ \epsilon \ 
\mathbb{Z}
.   \label{eq:11}
\end{equation}
and $t<T_{\infty }.$
Then the trajectory is periodic with the period.
\begin{equation}
T_{p}=\frac{\left( 4mK+2niK^{\prime }\right) {\small e}^{-i\pi /2}}{\sqrt{%
a\left( x_{0}-x_{2}\right) \left( x_{1}-x_{3}\right) }}.    \label{eq:12}
\end{equation}
Note that if $T_{p}>T_{\infty }$ the trajectory is still nonperiodic. By
imposing the condition that Im$\left( T_{p}\right) =0$, we have
\begin{equation}
r\equiv \frac{n}{m}=\frac{Im[2iK/z]}{Im\left[ K^{\prime }/z%
\right] };\ \ \ \ m,\ n\ \epsilon \ 
\mathbb{Z}
  \label{eq:13}
\end{equation}
where $z=\sqrt{a\left( x_{0}-x_{2}\right) \left( x_{1}-x_{3}\right) }.$
Since $n$ and $m$ are integers and the energy $E$ is fixed, $r$ is rational
and the equation (13) provides a discrete set of parameter values in the
complex $b$ plane for which classical trajectories are periodic. Let $%
b=b_{r}e^{i\theta }.$ Figures \ref{fig1}a and \ref{fig1}b show how the ratio $r$ varies with
discrete values of $\theta $ for the cases $k=1$ and $k=2$ respectively.
Without loss of generality, the energy $E$ is taken as unity as it is real.
The results can be generalized for any real energy $E$ by simple rescaling
of $x$ and $t$.

\section{Discretization of classical energy}\label{sec:3}

\begin{figure}[h]
\centering  
\includegraphics[width=7.5cm,height=6.0cm]{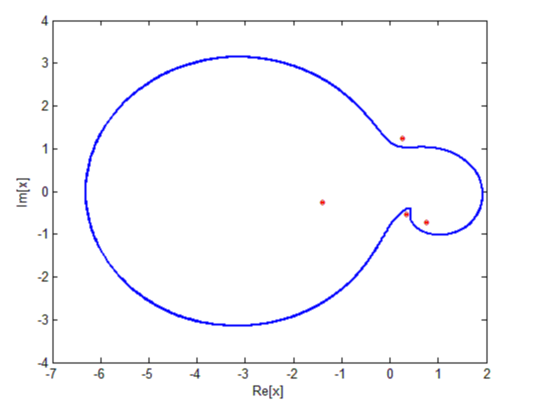}
\caption{\small{A periodic trajectory corresponding to $(m,n)=(1,2)$ for the
quartic potential $V(x)=x^{4}+(1+i)x$ with real energy $E=0.71624$}}
\label{P_1}
\end{figure}

\begin{figure}[h]
\centering  
\includegraphics[width=7.5cm,height=6.0cm]{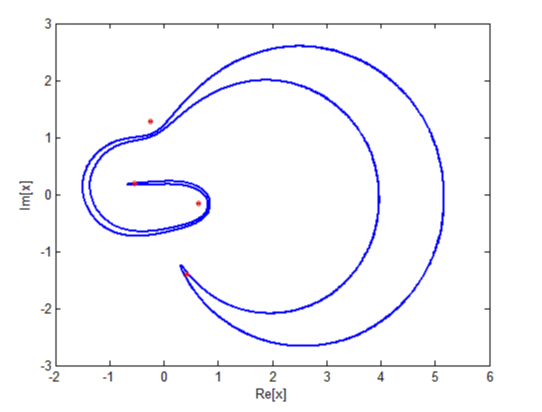}
\caption{\small{A periodic trajectory for the quartic potential $%
V(x)=x^{4}+(1+i)x^{2}$ corresponding to $(m,n)=(3,2)$ with real energy $%
E=0.81963$}}
\label{P_2}
\end{figure}

Next we consider the case when parameter $b$ is a fixed complex number and $%
E $ is a variable (Assume $a=1$ and $b=1+i.$). As a result, the equation
(\ref{eq:13}) allows only a discrete set of values of $E$ for which trajectories are
periodic. It was found that these discrete values of $E$ can be either real
or complex satisfying the condition (\ref{eq:13}). Table \ref{tab:Table_1} and Table \ref{tab:Table_2} show some real
discrete values of $E$ which make trajectories periodic when $k=1$ and $k=2$
respectively. Figures \ref{P_1} and \ref{P_2} show the periodic trajectories of systems $k=1$
and $k=2$ for two values of real energies.

\begin{figure}[h]
\centering  
\includegraphics[width=7.5cm,height=6.0cm]{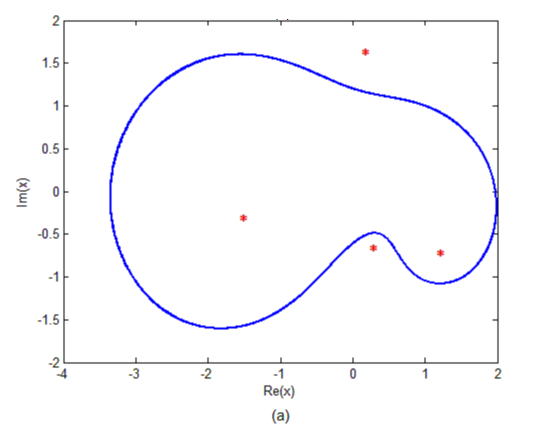}
\includegraphics[width=7.5cm,height=6.0cm]{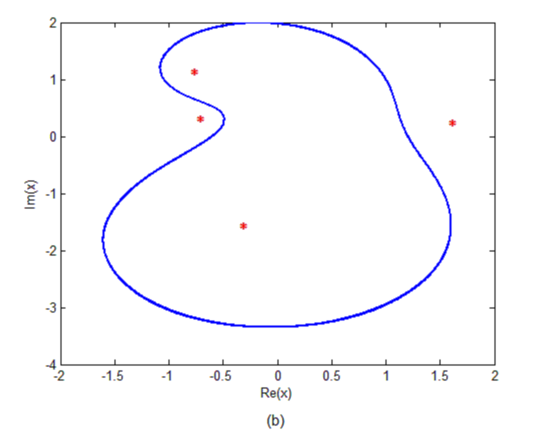} 
\caption{\small{A typical periodic classical trajectories of the potentials (a) $%
V(x)=x^{4}+(2+3i)x$ and (b) $V(x)=-x^{4}+(3+2i)x$\ . Energies of
trajectories\ corresponding to figures (a) and (b) are $E=2.42227$ and $%
E=-2.42227$ respectively. The four turning points are marked as dots.}}
\label{P_3}
\end{figure}

Moreover, it was found that if energy $E$ is corresponding to the periodic
trajectories of $p^{2}+ax^{4}+bx$ then $-E$ will be the energy which makes
trajectories of $p^{2}-ax^{4}+i\overline{b}x$ ( $\overline{b}$ is the
complex conjugate of $b$) periodic. Further $E$ and $-E$ are solutions
corresponding to the same $n$ and $m$ in the periodic condition (\ref{eq:13}) for
these two Hamiltonians respectively. In other words if $S_{E}$ is the
discrete set of energies for which classical trajectories of $%
p^{2}+ax^{4}+bx $ are periodic then $S_{-E}$ is the set of energies for
which trajectories of $p^{2}-ax^{4}+i\overline{b}x$ are periodic. Figures \ref{P_3}a and \ref{P_3}b show two periodic trajectories illustrating the above claim.

\begin{table}[h]
\centering
\begin{tabular}{ccc}
\hline
\textbf{m} &\hspace{1in}\ \textbf{n}\hspace{1in} & \textbf{E} \\
 \hline
1 & 1 & 0.27499 \\ 
1 & 2 & 0.71624 \\ 
1 & 3 & 0.78605 \\
2 & 3 & 0.60480 \\
2 & 5 & 0.74280 \\ 
2 & 1 & -0.28103 \\
3 & 1 & -0.53968 \\ 
3 & 2 & -0.07449 \\ 
5 & 2 & -0.42562 \\ 
\hline
\end{tabular}
\caption{ Classical energy spectrum corresponding to periodic trajectories of 
$V(x)=x^{4}+(1+i)x$ for various $(m,n).$}
\label{tab:Table_1}
\end{table}

\begin{table}[h]
\centering
\begin{tabular}{ccc}
\hline
\textbf{m} &\hspace{1in}\ \textbf{n}\hspace{1in} & \textbf{E} \\
 \hline
1 & 1 & -0.02143 \\ 
1 & 2 & -0.16951 \\ 
1 & 3 & -0.32417 \\ 
2 & 3 & -0.08940 \\ 
2 & 5 & -0.24827 \\ 
2 & 1 & 1.45802 \\ 
3 & 1 & 2.99725 \\ 
3 & 2 & 0.81963 \\ 
5 & 2 & 2.17849 \\ 
\hline
\end{tabular}
\caption{ Classical energy spectrum corresponding to periodic trajectories of 
$V(x)=x^{4}+(1+i)x^{2}$ for various $(m,n)$}
\label{tab:Table_2}
\end{table}

\section{\textbf{Periodic Classical trajectories of }$H=p^{2}+\protect\mu %
_{r}e^{i\protect\theta }x^{4}$}\label{sec:4}

Next we assume that $a=\mu $ and $b=0$ in the Hamiltonian (\ref{eq:1}). Then new
Hamiltonian has the form 
\begin{equation}
H=p^{2}+\mu x^{4}  \label{eq:14}
\end{equation}%
\ where $\mu $ is complex and $\mu =\mu _{r}e^{i\theta }$.
The equation of motion is
\begin{equation}
\frac{dx}{dt}=2\sqrt{E-\mu x^{4}}   \label{eq:15}
\end{equation}%
where $E$ is the total energy. Following the same procedure as in section \ref{sec:2},
we obtain required equations. By integrating (\ref{eq:15}) we have
\begin{equation}
\int \frac{dx}{\sqrt{E-\mu x^{4}}}=2t+c  \label{eq:16}
\end{equation}%
where $c$ is the constant of integration which depends on initial
conditions. The left-hand side of the above equation is an elliptic integral
of the first kind and hence equation (\ref{eq:16}) becomes
\begin{equation}
{\small F}\left( \sin ^{-1}\left[ \left( \frac{\mu }{E}\right) ^{1/4}x(t)%
\right] ,-1\right) {\small =2}\left( \mu E\right) ^{1/4}{\small t+\alpha } 
  \label{eq:17}
\end{equation}%

where $\alpha =\left( \mu E\right) ^{1/4}c$ and $F$\ is an elliptic
function. We invert the above equation in terms of Jacobian elliptic
function `$sn$' as
\begin{equation}
x\left( t\right) =\left( \frac{E}{\mu }\right) ^{1/4}sn\left( 2\left( \mu
E\right) ^{1/4}t+\alpha \ ;\ -1\right) .    \label{eq:18}
\end{equation}
Note that modulus $\kappa ^{2}=-1$ for the above problem. $K$ and $%
K^{\prime }$ are defined in (\ref{eq:7}) and (\ref{eq:8}) and
\begin{equation}
\kappa^{\prime^{2}}=1-\kappa ^{2}=2.    \label{eq:19}
\end{equation}
Then $K$ and $K^{\prime }$ are obtained as
\begin{equation}
K=\frac{\sqrt{\pi }\Gamma \left( 1/4\right) }{4\Gamma \left( 3/4\right) } 
  \label{eq:20}
\end{equation}
\begin{equation}
K^{\prime }=\frac{\sqrt{\pi }\Gamma \left( 1/4\right) }{4\Gamma \left(
3/4\right) }\left( 1-i\right)    \label{eq:21}
\end{equation}
As in the previous sections the condition of trajectory become
periodic and particle does not escape to infinity is
\begin{equation}
2\left( \mu E\right) ^{1/4}t=4mK+2niK^{\prime }\ ;\ \ \ m,\ n\ \epsilon \ 
\mathbb{Z}
.   \label{eq:22}
\end{equation}
Then the trajectory is periodic with the period
\begin{equation}
T_{p}(\mu _{r})=\frac{2mK+niK^{\prime }}{\left( \mu E\right) ^{1/4}}. 
  \label{eq:23}
\end{equation}
Since $\mu =\mu _{r}e^{i\theta }$
\begin{equation}
T_{p}(\mu _{r})=\frac{K}{\left( \mu _{r}E\right) ^{1/4}}\left[ \left(
2m+n\right) +in\right] \left( \cos \left( \theta /4\right) -i\sin \left(
\theta /4\right) \right) .   \label{eq:24}
\end{equation}
\begin{equation}
T_{p}(\mu _{r})=\frac{K}{\left( \mu _{r}E\right) ^{1/4}}\left[ \left( \left(
2m+n\right) \cos \left( \theta /4\right) +n\sin \left( \theta /4\right)
\right) +i\left( n\cos \left( \theta /4\right) -\left( 2m+n\right) \sin
\left( \theta /4\right) \right) \right] .    \label{eq:25}
\end{equation}
Since $K$ is \ real and $E$ is real and positive, by imposing the condition
that Im$\left( T_{p}\right) =0$, we have

\begin{equation}
\frac{m}{n}=\frac{\cot \left( \theta /4\right) -1}{2};\ \ \ \ m,\ n\
\epsilon \ 
\mathbb{Z}
.  \label{eq:26}
\end{equation}
or
\begin{equation}
\theta =4\tan ^{-1}\left[ \frac{n}{2m+n}\right] ;\ \ \ \ m,\ n\ \epsilon \ 
\mathbb{Z}
.   \label{eq:27}
\end{equation}
When $n=0$ and $m\neq 0$ , $H=p^{2}+\mu _{r}x^{4}$ and it is Hermitian. Then 
$H$ possesses periodic trajectories and the period $T_{p}(\mu _{r})$ becomes
 $T_{p}(\mu _{r})=\frac{2mK}{\left( \mu _{r}E\right) ^{1/4}}$ but
the period is corresponding to the minimum non zero $m\ $and the resulting
period is
\begin{equation}
T_{p+}(\mu _{r})=\frac{\sqrt{\pi }\Gamma \left( 1/4\right) }{2\left( \mu
_{r}E\right) ^{1/4}\Gamma \left( 3/4\right) }   \label{eq:28}
\end{equation}
On the other hand when $n\neq 0$ and $m=0$ ,$H=p^{2}-\mu _{r}x^{4}$ and it
is the non-Hermitian `wrong sign' potential which also possesses periodic
trajectories. The period is

\begin{equation}
T_{p-}(\mu _{r})=\frac{\sqrt{\pi }\Gamma \left( 1/4\right) }{2\sqrt{2}\left(
\mu _{r}E\right) ^{1/4}\Gamma \left( 3/4\right) }=\frac{\sqrt{\pi }\Gamma
\left( 1/4\right) }{2\left( 4\mu _{r}E\right) ^{1/4}\Gamma \left( 3/4\right) 
}    \label{eq:29}
\end{equation}
It is evident from equations (29) and (30) that the Hamiltonians $p^{2}+4\mu
_{r}x^{4}\ $\ and $p^{2}-\mu _{r}x^{4}$ have the same classical period (i.e. 
$T_{p+}(4\mu _{r})=T_{p-}(\mu _{r})$). Note that these two Hamiltonians are
the classical limit of the quantum mechanical isospectral Hamiltonians as
shown in \cite{R22, R23, R24, R25, R26}.

\bigskip 

\section{Concluding Remarks}\label{sec:5}

In this paper we have presented three main results. The first is that, for
given real energy, the classical trajectories of quartic Hamiltonians $%
H=p^{2}+ax^{4}+bx^{k}$, (where $a$ is real, $b$ is complex, and $k=1$ $or$ $%
2 $) are closed and periodic only for a discrete set of parameter curves in
the complex $b$-plane.

The second result is that given complex parameter $b$, the classical
trajectories are found to be periodic only for a discrete set of real
energies. As a result, real classical energies get discretized or quantized
by the condition that trajectories are periodic and closed. This result is
analogous to what was obtained by Anderson et al in \cite{R20} for real potential
parameters with complex $E$ (Here it is for complex potential parameters
with real energies). Further we showed that if $S(E)$ is the discrete set of
energies for which classical trajectories of $p^{2}+ax^{4}+bx$ are periodic
then $S(-E)$ is the set of energies for which trajectories of $p^{2}-ax^{4}+i%
\overline{b}x$ are periodic. We presented our results with illustrations. It
is important to note that when $b$ is complex and not pure imaginary, the
entire quantum eigen spectrum corresponding to the Hamiltonian $H$ is
complex and eigenenergies do not come as complex conjugate pairs. Therefore $%
H$ cannot be pseudo Hermitian and cannot have any antilinear symmetry.

As the third result, we showed that for real energies , the classical
trajectories of complex Hamiltonian $H=p^{2}+%
\mu
x^{4},$ ($%
\mu
=re^{i\theta })$ are periodic only for discrete values of $\theta $
satisfying the condition \ $\theta =4tan^{-1}[(n/(2m+n))]$ for $n$ and $m\in 
\mathbb{Z}
$. Further it was found that Hamiltonians $p^{2}+4\mu _{r}x^{4}$ $\ $\ and $%
p^{2}-\mu _{r}x^{4}$ which are the classical limit of the quantum mechanical
isospectral Hamiltonians introduced in \cite{R22,R23,R24,R25,R26}, have the same classical
period.

\end{document}